\documentclass[twocolumn,showpacs,preprintnumbers,amsmath]{revtex4}
\usepackage{dcolumn}
\usepackage{bm}
\begin{document}

\title{Non-Poisson processes: regression to equilibrium versus equilibrium correlation functions}
\author{Paolo Allegrini$^{1}$, Paolo Grigolini$^{2,3,4}$, Luigi Palatella$^{5}$, Angelo Rosa$^{6}$, Bruce J. West$^{7}$}
\address{$^{1}$ INFM, unit{\`a} di Como, via Valleggio 11, 22100 Como, Italy}
\address{$^{2}$Center for Nonlinear Science, University of North Texas,\\
P.O. Box 311427, Denton, Texas 76203-1427}
\address{$^{3}$Dipartimento di Fisica dell'Universit{\`a} di Pisa and INFM,\\
Via Buonarroti 2, 56127 Pisa, Italy}
\address{$^{4}$Istituto dei Processi Chimico Fisici del CNR, Area
della Ricerca di Pisa, Via G. Moruzzi 1, 56124 Pisa, Italy}
\address{$^{5}$ Dipartimento di Fisica and INFM, Center for Statistical Mechanics and Complexity,\\
Universit\`a di Roma "La Sapienza", P.le A. Moro 2, 00185 Rome, Italy}
\address{$^{6}$Institut de Math{\'e}matiques Bernoulli, Facult{\'e} 
des Sciences de Base, {\'E}cole Polytechique F{\'e}d{\'e}rale de 
Lausanne, 1015 Lausanne, Switzerland}
\address{$^{7}$Mathematics Division, Army Research Office, Research 
Triangle Park, NC 27709, USA}

\begin{abstract}
We study the response to perturbation of non-Poisson dichotomous fluctuations that generate super-diffusion.  We adopt the Liouville 
perspective and with it  a quantum-like approach based on splitting 
the density distribution into a symmetric and an anti-symmetric 
component. To fit the equilibrium condition behind the stationary 
correlation function, we study the time evolution of the 
anti-symmetric component, while keeping the symmetric component at 
equilibrium. For any realistic form of perturbed distribution density 
we expect a breakdown of the Onsager principle, namely, of  the 
property that
the subsequent regression of the perturbation to equilibrium is identical to
the corresponding equilibrium correlation function.  We find the 
directions to  follow for the calculation of higher-order 
correlation functions, an unsettled problem, which has been 
addressed in the past by means of approximations yielding quite 
different physical effects. 
\end{abstract}


\maketitle

\section{Introduction}\label{intro}

Complex physical systems typically have both nonlinear dynamical and
stochastic components, with neither one dominating. The response of such
systems to external perturbations determines the measurable characteristics
of phenomena from the beating of the human heart to the relaxation of
stressed polymers. What distinguishes such complex phenomena from processes
successfully studied using equilibrium statistical mechanics is how these
systems internalize and respond to environmental changes. Consequently, it
is of broad interest to determine which of the prescriptions from
equilibrium statistical physics is still applicable to complex dynamical
phenomena and which are not. Herein we address the breakdown of one of these
fundamental relations, that being, the Onsager Principle.  In the case of
ordinary statistical mechanics an exhaustive treatment of the relaxation of
perturbations to equilibrium can be found in Ref. \cite{kubobook}. Let us
consider as a prototype of ordinary statistical mechanics the case when the
stochastic variable under study $\xi(t)$ is described by the linear Langevin
equation
\begin{equation}
\frac{d\xi}{dt}=-\gamma \xi (t)+\eta(t),
\end{equation}
where the random driving force $\eta(t)$ is white noise. Let us imagine that
$\xi(t)$ is the velocity of a particle with unit mass and a given electrical
charge. Furthermore, we assume that this system reaches the condition of
equilibrium, and at a given time $t=0$, we apply an electrical field $E(t)$.
The external field $E(t)$ is an arbitrary function of time, fitting the
condition that $E(t)=0$, for $t<0$. The adoption of linear response theory
yields the prescription for the mean response of the system to the external
field
\begin{equation}
\langle \xi(t) \rangle = \int_{0}^{t}\Phi _{\xi}(t^{\prime 
})E(t-t^{\prime })dt^{\prime },
\end{equation}
where $\Phi _{\xi}(t)$ is the equilibrium correlation function of 
$\xi$. In this
familiar case, of dissipative Brownian motion, the autocorrelation function of the particle velocity is the exponential $\exp(-\gamma t)$.

There are two limiting cases of time dependence of the perturbation: (a) the
electric field is proportional to the Heaviside step function, $E(t)=K\Theta
(t)$, and is therefore a constant field after it is turned on at $t=0$; (b)
the electric field is proportional to the Dirac delta function, 
$E(t)=K\delta (t)$,
and is consequently an initial pulse that perturbs the
particle velocity. In these two limiting cases we obtain for the velocity of
the Brownian particle
\begin{equation}\label{greenkubo}
\langle \xi(t) \rangle = K\int_{0}^{t}\Phi _{\xi}(t^{\prime })dt^{\prime }
\end{equation}
and
\begin{equation}\label{onsager}
\langle \xi(t) \rangle = K\Phi _{\xi}(t),
\end{equation}
respectively. These two limiting cases show that in the case of ordinary
statistical mechanics the system's response to an external perturbation is
expressed in terms of the unperturbed autocorrelation function. We shall
refer to the conditions of Eq. (\ref{greenkubo}) and Eq. (\ref{onsager}) as
the Green-Kubo relation and the Onsager relation, respectively.

The search for a dynamical derivation of anomalous diffusion, that being
where the mean square value of the dynamic variable is not linear in time,
has been a subject of great interest in recent years.
There are two main theoretical perspectives on how to explain the 
origin of anomalous
diffusion. The first perspective is based on the assumption that there are
unpredictable events, that the occurrence of these events obey non-Poisson
statistics, and is related to the pioneering paper by Montroll and Weiss
\cite{montroll}.
The other perspective rests on the assumption that the
single diffusion trajectories have an infinite memory. The prototype of the
latter perspective is the concept of fractional Brownian motion
introduced by Mandelbrot \cite{mandelbrot}.
A problem worthy of investigation is as to whether or not, in the case of anomalous diffusion,
the response to external perturbation departs from the predictions of Eqs. (%
\ref{greenkubo}) and (\ref{onsager}). 
In the last few years, this problem has been addressed by some investigators \cite{elena,barkai1,barkai2,barkaifleurov,metzlerklaftersokolov,sokolovblumenklafter}.
These authors have discussed the Green-Kubo relation of Eq. (\ref{greenkubo}). Notice that in the special case of ordinary statistical
mechanics this relation can also be written in the following form
\begin{equation}\label{generalizedeinsteinrelation}
\langle x(t) \rangle = \frac{K}{2 \langle {\xi}^{2} \rangle} \langle
x^{2}(t) \rangle_{0}.
\end{equation}
To understand how to derive this equation, originally proposed by Bouchaud
and George \cite{bouchaud}, we have to refer ourselves to the following equation of motion
\begin{equation}\label{diffusion}
\frac{dx}{dt}=\xi(t).
\end{equation}
Since $\xi(t)$ is a fluctuating velocity, it generates spatial
diffusion and we denote by $x(t)$ the position of the corresponding
diffusing particle. The external field affects the velocity fluctuation and, consequently, the diffusion process
generated by these fluctuations. In the absence of perturbation, the second
moment of the diffusing particle, $\langle x(t)^{2} \rangle_{0}$, obeys the prescription
\begin{equation}\label{secondmoment}
\langle x(t)^{2} \rangle_{0}=2 \langle {\xi}^{2} \rangle
\int_{0}^{t}dt^{\prime }\int_{0}^{t^{\prime}}dt^{\prime \prime }\Phi _{\xi}(t^{\prime \prime }).
\end{equation}
It is straightforward to prove that Eq. (\ref{greenkubo}) yields Eq. (\ref
{generalizedeinsteinrelation}). This is done by considering the mean 
value of 
Eq.(\ref{diffusion})
\begin{equation}
\label{intermediatestep}
\frac{d\langle x \rangle}{dt} = \langle \xi(t) \rangle.
\end{equation}
The time integration of the left hand term of this equation yields the 
left hand term of Eq. (\ref{generalizedeinsteinrelation}) and the 
time integration of the right hand term of it, using 
Eq.(\ref{greenkubo}) and Eq.(\ref{secondmoment}), yields the right 
hand term of Eq. (\ref{generalizedeinsteinrelation}).

The relation of Eq. (\ref{generalizedeinsteinrelation}) is denoted as 
\emph{generalized Einstein relation}, because it might hold true also 
when the equilibrium correlation function does not exist \cite{barkaifleurov}.
However, in the case of ordinary statistical mechanics Eq. (\ref{generalizedeinsteinrelation})
becomes equivalent to the Green-Kubo
property. In this generalized sense we can state that the authors of Refs.
\cite{elena,barkai1,barkai2,barkaifleurov,metzlerklaftersokolov,sokolovblumenklafter}
studied the Green-Kubo relation of Eq. (\ref{greenkubo}), all of them but
the authors of Ref. \cite{elena}, devoting their attention to the
subdiffusional case.

Herein we focus our attention on the Onsager relation of Eq. (\ref{onsager}).
The only earlier work on the issue of the breakdown of the Onsager
principle, caused by anomalous statistics, known to us, is that of 
Ref. \cite{gerardo}.
However, here we plan to address the problem with the adoption of
a Liouville-like approach, a fact that will allow us to establish some
general conclusions concerning the breakdown of the Onsager prescription.
We shall illustrate the rules for the calculation of the 
four-time correlation, with a prescription that can be easily 
extended to correlation  functions of any order. We expect that these
prescriptions might lead to a successful evaluation of the 
fourth-order correlation function, which has been studied so far by 
means of a factorization assumption which is violated by the 
non-Poisson statistics.

\section{An idealized model of intermittent randomness and the corresponding density equation}\label{themodel}
As done in Ref. \cite{maurobologna}, let us focus on the following dynamical
system. Let us consider a variable $y$ moving within the interval $I=[0,2]$.
The interval is defined over an overdamped potential $V,$ with a cusp-like
minimum located at $y=1$. If the initial condition of the particle is $y(0)>1
$, the particle moves from the right to the left towards the potential
minimum. If the initial condition is $y(0)<1$, then the motion of the
particle towards the potential minimum takes place from the left to the
right. When the particle reaches the potential bottom it is injected to an
initial condition, different from $y=1$, chosen in a random manner. We thus
realize a mixture of randomness and slow deterministic dynamics. The left
and right portions of the potential $V(y)$ correspond to the laminar regions
of turbulent dynamics, while randomness is concentrated at $y=1$. In other
words, this is an idealization of the map used by Zumofen and Klafter \cite
{klafter1}, which does not affect the long-time dynamics of the process,
yielding only the benefit of a clear distinction between random and
deterministic dynamics. Note that the waiting time distribution in the two
laminar phases of the reduced form has the same time asymptotic form as
\begin{equation}\label{waiting}
\psi(t)=(\mu -1)\frac{T^{\mu -1}}{(t+T)^{\mu }}.
\end{equation}
We select this form as the simplest possible way to ensure the normalization
condition
\begin{equation}\label{normalization}
\int_{0}^{\infty }dt\psi (t)=1.
\end{equation}
We note that Eq. (\ref{normalization}) implies $\mu >1$. The condition $\mu >2$ 
corresponds to the existence of a finite mean sojourn time, and, thus, to
the possibility itself of defining the stationary correlation function of
the fluctuation $\xi$, which, with the choice of Eq. (\ref{waiting})
reads \cite{geisel3}
\begin{equation}\label{correlationfunction}
\Phi _{\xi}(t)\equiv \frac{\langle \xi(t)\xi(0) \rangle}{\langle{\xi}^{2} \rangle} = \left[\frac{T}{t+T}\right]^{\mu -2}.
\end{equation}

From within the perspective of a single trajectory this dynamical model reads
\begin{equation}\label{new_doty}
\dot{y}=\lambda [\Theta (1-y)y^{z}-\Theta(y-1)(2-y)^{z}]+\frac{\Delta_{y}(t)}{\tau _{random}}\delta (y-1).
\end{equation}
The function $\Theta (x)$ is the ordinary Heaviside step function, 
$\Delta_{y}(t)$ is a random function of time that can achieve any 
value on the interval $[-1,+1]$, and $\tau _{random}$ is the injection time that must
fulfill the condition of being infinitely smaller than the time of
sojourn in one of the laminar phases. Note that $z$ is a real number fitting
the condition $z>1$. In fact, the equality
\begin{equation}
z=\frac{\mu }{(\mu -1)}
\end{equation}
relates the dynamics of Eq. (\ref{new_doty}) to the distribution Eq. (\ref{waiting}). The Poisson condition is recovered in the limit 
$z\rightarrow 1$, namely, in the limit $\mu \rightarrow \infty $.
Thus, in a sense, the whole region $z>1$ ($\mu <\infty $) corresponds to anomalous statistical
mechanics. However, the deviation from normal statistical mechanics is
especially evident when $z>1.5$, a condition implying that the second moment
diverges. In the case $z>2$ the departure from ordinary statistical
mechanics becomes even more dramatic, due to the fact that the first moment
also diverges and, as we shall see in this Section, the process becomes
non-ergodic.

Let us move now to the density picture, namely, to a formulation of 
Eq. (\ref{new_doty}) from within the Gibbs perspective.
The form of this equation is:
\begin{equation}\label{evol_operator}
\frac{\partial }{\partial t}p(y,t)=-\lambda \frac{\partial }{\partial 
y}[\Theta (1-y)y^{z}-\Theta (y-1)(2-y)^{z}]p(y,t)+C(t),
\end{equation}
where
\begin{equation}\label{obviousformforinjectionback}
C(t)=2\lambda p(1,t).
\end{equation}
It is important to stress that we are forced to set the equality of 
Eq. (\ref{obviousformforinjectionback}) to fulfill the following 
physical conditions
\begin{equation}\label{define_ct}
\frac{d}{dt}\int_{I=[0,2]}p(y,t)dy=\int_{I=[0,2]}\frac{\partial 
}{\partial t}p(y,t)dy=0,
\end{equation}
which, in turn, ensures the conservation of probability. We assume the
ordinary normalization condition
\begin{equation}\label{normalizationcondition}
\int_{I=[0,2]}p(y,t)dy=1,
\end{equation}
which is kept constant in time, as a consequence of Eq. (\ref{define_ct}). It
is evident that the inhomogeneous term $C(t)$ corresponds to the action of
the stochastic term, namely, the second term on the right hand side 
of Eq. (\ref{new_doty}).

It is important to point out that our dynamic perspective allowed us to
describe the intermittent process through the Liouville-like equation
\begin{equation}\label{idealization}
\frac{\partial p(\xi,y,t)}{\partial t}=\mathcal{R}p(\xi,y,t),
\end{equation}
where $y$ denotes a continuous variable moving either in the right or in the
left laminar region, with $\xi$ getting the values $W$ or $-W$,
correspondingly, and the operator $\mathcal{R}$ reading
\begin{equation}\label{densityoperator}
\mathcal{R}=-\lambda \frac{\partial }{\partial y}[\Theta 
(1-y)y^{z}-\Theta(y-1)(2-y)^{z}]+2\lambda \int_{0}^{2}dy\delta (y-1).
\end{equation}
This operator departs from the conventional form of a differential operator,
since the last term corresponds to the unusual role of an injection process,
which is random rather than being deterministic. In the ordinary
Fokker-Planck approach the role of the stochastic force is played by a
second-order derivative, which is not as unusual as the integral operator of
Eq. (\ref{densityoperator}). Here the role of randomness is played by the
back-injection process, which, from within the density perspective is
described by an operator that selects from all possible values $p(y,t)$, the
specific value of $p(y,t)$ at $y=1$. The idealization that we have adopted,
of reducing the size of the chaotic region to zero, with the choice of the
process of back injection located at $y=1$, makes it possible for us to use
the continuous time representation and the equation of motion Eq. 
(\ref{idealization}) rather than the conventional Frobenius-Perron
representation. This representation will allow us to obtain analytical
results. However, the same physical conclusions would be reached, albeit
with more extensive algebra, using the conventional maps and the
Frobenius-Perron procedure described in the recent book by Driebe \cite
{driebe}.

We note that the equilibrium probability density solving Eq. (\ref
{evol_operator}) is given by
\begin{equation}\label{inv_distr}
p_{0}(y)=\frac{2-z}{2}\left[ \frac{\Theta 
(1-y)}{y^{z-1}}+\frac{\Theta (y-1)}{(2-y)^{z-1}}\right].
\end{equation}
This equilibrium density becomes negative for $z>2$ and signals the
important fact that for $z>2$ there no longer exists an invariant
distribution. The lack of an invariant distribution accounts for the
nonergodicity in the fluorescence of single nanocrystals, recently pointed
out by Brokmann \emph{et al.} \cite{nonergodicity}.

For the purposes of calculation in the next few Sections, it is convenient
to split the density $p(y,t)$ into a symmetric and an anti-symmetric part
with respect to $y=1$,
\begin{equation}\label{anti-symm1}
p(y,t)=p_{S}(y,t)+p_{A}(y,t).
\end{equation}
This separation based on symmetry yields the following two equations 
from Eq. (\ref{evol_operator})
\begin{eqnarray}\label{new_evol_symm}
\frac{\partial}{\partial t}p_S(y, t) & = & - \lambda \Theta(1-y) 
\frac{\partial}{\partial y} [y^z p_S(y, t)]  \nonumber\\
& & + \lambda \Theta(y-1) \frac{\partial}{\partial y}[(2-y)^z p_S(y, t)] +
C(t)  \nonumber \\
\end{eqnarray}
and
\begin{eqnarray}\label{new_evol_anti}
\frac{\partial}{\partial t}p_A(y, t) & = & - \lambda \Theta(1-y) 
\frac{\partial}{\partial y} [y^z p_A(y, t)]  \nonumber\\
& & + \lambda \Theta(y-1) \frac{\partial}{\partial y}[(2-y)^z p_A(y, t)] .
\end{eqnarray}

We note that the anti-symmetric part of the density is driven by a
conventional differential operator, which we denote by $\hat{\Gamma}$. Thus,
we rewrite Eq. (\ref{new_evol_anti}) as follows
\begin{equation}\label{ordinary}
\frac{\partial }{\partial t}p_{A}(y,t)=\hat{\Gamma}p_{A}(y,t),
\end{equation}
where
\begin{equation}\label{usualoperator}
\hat{\Gamma}\equiv -\lambda \Theta (1-y)\frac{\partial }{\partial 
y}y^{z}+\lambda \Theta (y-1)\frac{\partial }{\partial y}(2-y)^{z}.
\end{equation}
The operator with the unusual form, containing $C(t)$, is only responsible
for the time evolution of the symmetric part of the probability density. We
notice, on the other hand, that any physical effect producing a departure of
$C(t)$ from its equilibrium value, if this exists, namely, if $z<2$,
implies a departure from equilibrium. A stationary correlation function can
be evaluated, as we shall see in the next few Sections, using only 
Eq. (\ref{new_evol_anti}),
without forcing Eq. (\ref{new_evol_symm}) to depart from
the equilibrium condition. 

As we shall see in Section \ref{higherorder}, the evaluation of correlation functions of order higher than the second cannot be done without producing a departure of $C(t)$ from its equilibrium value. 
This might generate the impression that the correlation functions of order higher than the second cannot be
evaluated without internal inconsistencies, if we use only the density
picture. The evaluation of these higher-order correlation functions was done  in Ref. \cite{allegro}, by using a procedure based on the time evolution of single trajectories.
Actually, as we shall see in Section \ref{higherorder}, the density approach should
yield  the same result.   However, we think that deriving this result using only the Liouville-like equation of this section is a hard task, which was bypassed in the past by means of the factorization approximation \cite{maurobologna},  violated by the non-Poisson case. 

\section{The correlation function of the dichotomous fluctuation from the trajectory picture}\label{trajpict}
Let us focus our attention on Eq. (\ref{new_doty}) and consider the initial
condition $y_{0}\in [0,1]$. Then, it is straightforward to prove that the solution, for $y<1$ is,
\begin{equation}\label{solution}
y(t)=y_{0}\left( 1-\lambda (z-1)y_{0}^{z-1}t\right) ^{-1/(z-1)}.
\end{equation}
From (\ref{solution}) and imposing the condition $y(T)=1$, we can find the
time at which the trajectory reaches the point $y=1$, which is
\begin{equation}\label{arrive}
T = T(y_{0}) = \frac{1-y_{0}^{z-1}}{\lambda (z-1)y_{0}^{z-1}}.
\end{equation}
Since we have to find $\xi(t)$ and $\xi(t)=\xi(y(t))$, from the general form of Eq. (\ref{arrive}) we obtain:
\begin{eqnarray}\label{xi_t}
\frac{\xi(t)}{W} &=&[\Theta (1-y_{0})\Theta (T(y_{0})-t) \nonumber\\
& & -\Theta (y_{0}-1)\Theta (T(2-y_{0})-t)]  \nonumber\\
& & -\sum_{i=0}^{+\infty }\mbox{sign}\left[ \Delta_{y}\left(\sum_{k=0}^{i}\tau _{k}\right) \right]\times \nonumber\\
& & \times \left[ \Theta \left( \sum_{k=0}^{i+1}\tau _{k}-t\right) - \Theta\left(\sum_{k=0}^{i}\tau _{k}-t\right) \right], \nonumber\\
\end{eqnarray}
where the time increments are given by
\begin{eqnarray}\label{define_tau}
\tau_{0} & = & T(y_{0})=\frac{1-y_{0}^{z-1}}{\lambda(z-1)y_{0}^{z-1}} \Theta(1-y_{0}) \nonumber\\
& & +\frac{1-(2-y_{0})^{z-1}}{\lambda (z-1)(2-y_{0})^{z-1}}\Theta(y_{0}-1) \nonumber\\
& & \nonumber\\
\tau _{i\ge 1} & = &\frac{1-[1+\Delta _{y}(\tau_{i})]^{z-1}}{\lambda(z-1)[1+\Delta _{y}(\tau _{i})]^{z-1}}\Theta(-\Delta _{y}(\tau _{i}))\nonumber\\
& & +\frac{1-[1-\Delta _{y}(\tau _{i})]^{z-1}}{\lambda (z-1)[1-\Delta_{y}(\tau _{i})]^{z-1}}\Theta (\Delta _{y}(\tau _{i})). \nonumber\\
\end{eqnarray}
Then, for the autocorrelation function we obtain the following expression:
\begin{eqnarray}\label{xit_xi0}
\frac{\langle \xi(t)\xi(0)\rangle}{W^{2}} & = & \langle [\Theta(1-y_{0})\Theta (T(y_{0})-t)  \nonumber\\
& & +\Theta (y_{0}-1)\Theta (T(2-y_{0})-t)]\rangle +  \nonumber\\
& & \sum_{i=0}^{+\infty }\left\langle\mbox{sign}(y_{0}-1)\mbox{sign}\left[\Delta _{y}\left(\sum_{k=0}^{i}\tau _{k}\right) \right] \times \right.\nonumber\\
& & \times \left. \left[ \Theta \left( \sum_{k=0}^{i+1}\tau _{k}-t\right)-\Theta \left( \sum_{k=0}^{i}\tau _{k}-t\right) \right] \right\rangle. \nonumber\\
\end{eqnarray}
As pointed out in Section \ref{themodel}, the calculation of the
correlation function rests on averaging on the invariant distribution given by Eq. (\ref{inv_distr}).
As a consequence of this averaging, the second term in (\ref{xit_xi0})
vanishes. In fact, the quantity to average is anti-symmetric, whereas the statistical weight is symmetric.

It is possible to write the surviving term in the autocorrelation as
\begin{eqnarray}\label{xit_xi0_bis}
\frac{\langle \xi(t)\xi(0) \rangle }{W^{2}} & = &(2-z)\int_{0}^{1}\Theta \left( \frac{1-y^{z-1}}{\lambda(z-1)y^{z-1}}-t\right) \frac{1}{y^{z-1}}dy  \nonumber\\
& = & (2-z)\int_{0}^{(1+\lambda (z-1)t)^{-1/(z-1)}}y^{-z+1}dy  \nonumber \\
& = & (1+\lambda (z-1)t)^{-(2-z)/(z-1)}  \nonumber \\
&\equiv & (1+\lambda (z-1)t)^{-\beta },
\end{eqnarray}
with
\begin{equation}\label{beta}
\beta =\frac{2-z}{z-1}.
\end{equation}
Since we focus our attention on $0<\beta <1$, we have to consider $3/2<z<2$.
Note that the region $1<z<3/2$ does not produce evident signs of deviation
from ordinary statistics. However, as we shall see in Section \ref{onsregr}, an exact agreement between density and trajectory is recovered only at $z=1$, when
the correlation function becomes identical to the exponential 
function $\exp(-\lambda t)$. Note also that Eq. (\ref{xit_xi0_bis}) becomes identical
to Eq. (\ref{correlationfunction}) after setting the condition
\begin{equation}\label{constraint}
\lambda (z-1)=\frac{1}{T}.
\end{equation}

\section{The correlation function of the dichotomous fluctuation from the density picture}\label{denspict}
The result of the preceding Section is reassuring, since it establishes that
the intermittent model we are using generates the wanted inverse power law
form for the correlation function of the dichotomous variable $\xi(t)$. In
this Section we show that exactly the same result can be derived from the
adoption of the Frobenius-Perron form of Eq. (\ref{evol_operator}).

To fix ideas, let us consider the following system: a particle in the
interval $[0,1]$ moves towards $y=1$ following the prescription $\dot{y} = \lambda y^{z}$ and when it reaches $y=1$ it is injected backwards at a
random position in the interval. The evolution equation obeyed by the
densities defined on this interval is the same as Eq. (\ref{evol_operator}),
with $C(t)=\lambda p(1,t)$. This dynamic problem was already addressed in
Refs. \cite{markos,massi}, and solved using the method of characteristics as
detailed in Ref. \cite{goldenfeld}. It is important to stress that this
approach is the requisite price for adopting the idealized version of
intermittency. The adoption of the more conventional reduced map of 
Ref. \cite{klafter1} would have made it possible for us to adopt the 
elegant prescriptions of Driebe \cite{driebe}, as done in Ref. \cite{markos}.

Let us remind the reader that the solution afforded by the method of
characteristics, in the case of this simple nonlinear equation with
stochastic boundary conditions is
\begin{eqnarray}\label{first_solution}
p(y,t) & = &\int_{0}^{t}\frac{\lambda p(1,\xi )}{g_{z}((t-\xi )y^{z-1})}d\xi \nonumber\\
& & + p\left(\left[\frac{y}{[g_{z}(y^{z-1}t)]^{1/z}}\right] ,0\right) \times \nonumber\\
& & \times \frac{1}{g_{z}(y^{z-1}t)}
\end{eqnarray}
where
\begin{equation}
g_{z}(x)\equiv (1+\lambda (z-1)x)^{z/(z-1)}.
\end{equation}

To find the autocorrelation function $\langle \xi(t)\xi(0)\rangle $ 
using only densities, we have to solve Eqs. (\ref{new_evol_symm}) and (\ref{new_evol_anti}), which are equations of
the same form as that yielding Eq. (\ref{first_solution}). For this reason we
adopt the method of characteristics again. To do the calculation in this
case, it is convenient to adopt a frame symmetric with respect to $y=1$.
Then, let us define $Y=y-1$. Using the new variable and Eq. (\ref{first_solution}), 
we find for Eqs. (\ref{new_evol_symm}) and (\ref{new_evol_anti}) the following solution
\begin{eqnarray}\label{solution1}
p_{S}(Y,t) & = & \int_{0}^{t}\frac{\lambda p_{S}(0,\tau)}{g_{z}((t-\tau )(1-|Y|)^{z-1})}d\tau + \nonumber\\
& & p_{S}\left(\left[1-\frac{1-|Y|}{[g_{z}((1-|Y|)^{z-1}t)]^{1/z}}\right] ,0\right) \times  \nonumber\\
& & \times \frac{1}{g_{z}((1-|Y|)^{z-1}t)}
\end{eqnarray}
and
\begin{eqnarray}\label{solution2}
p_{A}(Y,t) & = & p_{A}\left( \mbox{sign}(Y)\left[1-\frac{1-|Y|}{[g_{z}((1-|Y|)^{z-1}t)]^{1/z}}\right] ,0\right) \times \nonumber\\
& & \times \frac{1}{g((1-|Y|)^{z-1}t)}.
\end{eqnarray}
Then, the solution consists of two terms: (1) the former is an \textit{even}
term and is responsible for the long time limit of the distribution
evolution and (2) the latter is an \textit{odd} term which disappears in the
long-time limit. We note that (2) is a desirable property because Eq. (\ref{new_evol_anti}) does not contain the injection term $C(t)$ and the
equilibrium density (\ref{inv_distr}) is an \textit{even} function,
independently of the symmetry of the initial distribution.

As pointed out in Section \ref{themodel}, the two-time correlation 
function is determined by the anti-symmetric part of the probability density alone. Thus, the definition
of the autocorrelation function does not conflict with the equilibrium
assumption. In Section \ref{onsregr}, we shall see that this conflict 
reemerges when we
attempt to evaluate higher-order correlation functions. Let us calculate the
autocorrelation function explicitly:
\begin{eqnarray}\label{correlation}
\langle \xi(t)\xi(0)\rangle
& = &(2-z)\left. \int_{0}^{1}\frac{1}{(1-\tau )^{z-1}}\right| _{\tau = 1-\frac{1-Y}{[g_{z}((1-Y)^{z-1}t)]^{1/z}}}\times \nonumber\\
& & \times \frac{1}{g_{z}((1-Y)^{z-1}t)}dY
\end{eqnarray}
The integral (\ref{correlation}) is exactly solvable and leads to the expression
\begin{equation}\label{correl}
\langle \xi(t)\xi(0) \rangle = (1+\lambda(z-1)t)^{-(2-z)/(z-1)},
\end{equation}
which is the same result as that found in Section \ref{trajpict}, using trajectories
rather than the probability densities.

In a similar way, it is possible to calculate the autocorrelation $\langle
Y(t)Y(0)\rangle $ and determine that its temporal behavior is an inverse
power law with the same exponent as that in Eq. (\ref{correl}).

\section{Onsager regression to equilibrium}\label{onsregr}
In conclusion, in the two preceding Sections we have established that the
Liouville-like representation of Eq. (\ref{idealization}) yields, as
expected, the equilibrium correlation function of Eq. 
(\ref{correlationfunction}) with
\begin{equation}
T\equiv \frac{\mu -1}{\lambda }.
\end{equation}

What about Onsager's regression to equilibrium? Let us first of all discuss
a physical condition where the prescription of Eq. (\ref{onsager}) is
fulfilled. Let us consider the initial distribution $p(y,0)$ defined as
follows
\begin{equation}
p(y,0)=0,\text{ }y<1
\end{equation}
and
\begin{equation}
p(y,0) = 2p_{0}\left( y\right), \text{ } y>1,
\end{equation}
where $p_{0}(y)$ denotes the equilibrium distribution of Eq. (\ref{inv_distr}).
It is convenient to point that this equilibrium distribution is symmetric
and that the factor of $2$ serves the purpose of normalizing the out-of
equilibrium condition that we are studying.
Let us denote by $\Delta P(t)$ the population difference between the 
left and the right state. The choice
we made sets the initial condition $\Delta P(0)=1$.
It is straightforward to show that Onsager's regression implies that 
$\Delta P(t)=\Phi _{\xi}(t)$.
Let us split the initial distribution in the symmetric and anti-symmetric part,
\begin{equation}\label{splitting}
p(y,0)=p_{0}(y)+p_{A}^{(eq)}(y,0).
\end{equation}
We note that
\begin{equation}
p_{A}^{(eq)}(y,0) = -p_{0}(y),y<1
\end{equation}
and
\begin{equation}
p_{A}^{(eq)}(y,0) = p_{0}(y),y>1.
\end{equation}
The symmetric part does not contribute to $\Delta P(t)$, only the 
anti-symmetric part does. Thus, we obtain
\begin{equation}\label{regression1}
\Delta P(t) = 2\int_{0}^{1}\exp(\hat{\Gamma}t)p_{A}^{(eq)}(y,0),
\end{equation}
where $\hat{\Gamma}$ is the operator driving the anti-symmetric part, defined
by Eq. (\ref{usualoperator}). With the choice of initial condition made it
is straightforward to prove that the right end side of Eq. (\ref{regression1})
is the equilibrium correlation function evaluated in Sections \ref{trajpict} and \ref{denspict}. Thus,
the prescription of ordinary statistical physics is fulfilled.

What about the regression to equilibrium in general? We note that we can
adapt the earlier arguments to any initial condition thereby yielding
\begin{equation}\label{regression2}
\Delta P(t)=2\int_{0}^{1}\exp(\hat{\Gamma}t)p_{A}^{(noneq)}(y,0).
\end{equation}
Note that $p_{A}^{(noneq)}(y,t)$ obeys the Liouville-like prescription
corresponding to $\frac{dy}{dt} = \lambda y^{z}$, namely, the equation of
motion of the left laminar region, without any back-injection process.
Since, as we have seen in Section \ref{themodel}, the back-injection 
term serves the
purpose of keeping constant the population of the system, we have 
that $\Delta P(t)\rightarrow 0$ for $t\rightarrow \infty $. More 
precisely, we
obtain
\begin{equation}\label{regressiontoequilibrium}
\frac{d\Delta P(t)}{dt} = 
2\hat{\Gamma}\int_{0}^{1}\exp(\hat{\Gamma}t)p_{A}^{(noneq)}(y,0) = 
-2p_{A}^{(noneq)}(1,t).
\end{equation}

The superscript $noneq$ serves the purpose of pointing out that in general
the perturbation process creating the necessary initial asymmetry 
will not realize, for the left portion of the asymmetric component, a 
form
exactly identical to the left portion of the equilibrium distribution. This
observation makes it possible to estimate the time asymptotic behavior of $%
\Delta P(t)$ in general. The exact time evolution of $\Delta P(t)$ depends
on the detailed effects of the perturbation. However, for any realistic
perturbation, we can prove that the time asymptotic behavior obeys an
universal prescription. If the perturbation does not affect the equilibrium
distribution in the regions close to the borders, namely for $y\leq \epsilon
$ and $y\geq 2-\epsilon $, where $\epsilon \ll 1$, we have that $p_{A}^{(noneq)}(y,0)$ vanishes for $y\leq \epsilon $. To evaluate the
asymptotic behavior of $\Delta P(t)$ we select the infinitesimal portion $%
dy(0)$ of the interval $[0,1]$, closest to $y=0$, where $p_{A}^{(noneq)}(y,0)
$ does not vanish. We call $M$ the number of trajectories located in this
interval at $t=0$. The asymptotic behavior of $\Delta P(t)$ is determined by
the time necessary for these trajectories to reach the border $y=1$. The
first trajectory will reach the border after a given time $t=T_{first}$,
after which the number $M$ will begin decreasing, thereby determining the
decay of $\Delta P(t)$ in this time asymptotic region. It is straightforward
to prove that the time of arrival at $y=1$ of the trajectory with initial
condition $y(0)$, called $t$, is related to $y(0)$ by
\begin{equation}
y(0)=\frac{1}{[1+(z-1)\lambda t]^{\frac{1}{z-1}}}.
\end{equation}
Thus we obtain
\begin{equation}
\frac{d\Delta P(t)}{dt}\approx \frac{dM}{dt}=\frac{1}{[1+(z-1)\lambda t]^{\frac{z}{z-1}}}.
\end{equation}
By integrating this equation and taking into account that $z=\mu /(\mu -1)$,
we finally obtain
\begin{equation}
\lim_{t\rightarrow \infty }\Delta P(t)\propto \frac{1}{t^{\mu -1}},
\label{onsagerbreakdown}
\end{equation}
which sanctions the breakdown of Eq. (\ref{onsager}).

\section{Higher-order correlation functions}\label{higherorder}
We now show
that the calculation of higher-order correlation functions, though 
difficult, can be done by 
establishing an even deeper connection 
with the quantum mechanical perspective. Let us address the problem 
of evaluating the fourth-order correlation
function $\langle \xi(t_{4})\xi(t_{3})\xi(t_{2})\xi(t_{1}) \rangle$.
According to the prescription
that we adopted in Section \ref{denspict} we must proceed as follows. 
We move from the
equilibrium distribution and let it evolve for a time $t_{1}$. The
distribution selected is in equilibrium. Therefore it will remain unchanged.
At time $t=t_{1}$ we apply the operator $\xi$ to the
distribution. Since the equilibrium distribution is symmetric, the
application of the sign operator changes it into the anti-symmetric
distribution. We let this distribution evolve for the time $t_{2}-t_{1}$.
This has the effect of yielding 
$$
\Phi_{\xi}(t_{2}-t_{1})p_{A}^{(eq)}(y,0)+p_{A}^{(noneq)}(y,t_{2}-t_{1}).
$$

At this stage there are two possibilities:
%\begin{description}
\begin{itemize}
\item[(a)] $p_{A}^{(noneq)}(y,t_{2}-t_{1})=0$; 
\item[(b)] $p_{A}^{(noneq)}(y,t_{2}-t_{1})\neq 0$.
\end{itemize}
%\end{description}

Let us consider the case (a) first. In this case, we proceed as follows. At
time $t_{2}$ we apply to the distribution the operator $\xi$, and we 
change it into the original equilibrium distribution. This means
that the time evolution from $t_{2}$ to $t_{3}$ leaves it unchanged. At time
$t_{3}$ we apply to it the operator $\xi$ and we turn it
into $p_{A}^{(eq)}$ again. We let this distribution evolve till to 
time $t_{4}$.
At this time we apply to it the operator $\xi$
again and we make the final average. The result of the condition (a) yields
\begin{equation}\label{factorization}
\langle \xi(t_{4})\xi(t_{3})\xi(t_{2})\xi(t_{1}) \rangle = \langle 
\xi(t_{4})\xi(t_{3}) \rangle \langle \xi(t_{2})\xi(t_{1}) \rangle.
\end{equation}
In a recent work \cite{allegro} it has been shown that this factorization
condition is violated by non-Poisson statistics. The demonstration was made
by applying the method of conditional probabilities to the study of single
trajectories. Thus we are forced to consider case (b).

The problem with condition (b) is that it yields a distribution with the
symmetric component departing from equilibrium. This departure is in 
an apparent conflict
with the assumption that the autocorrelation function is calculated
using the equilibrium condition. In fact it seems to be  equivalent 
to stating that the calculation of an equilibrium correlation 
function generates an out of equilibrium condition.  Let us see why. 

In the case of the two-time correlation function we apply the 
operator $\xi$ to the density distribution twice. The first 
application allows us to observe the time evolution of the 
anti-symmetric component distribution density, with no conflict with 
equilibrium, given the fact  that the symmetric component remains at 
equilibrium. 
The second application of $\xi$  turns 
$$
\Phi_{\xi}(t_{2}-t_{1})p_{A}^{(eq)}(y,0)+p_{A}^{(noneq)}(y,t_{2}-t_{1})
$$ 
into 
$$
\Phi_{\xi}(t_{2}-t_{1})p_{S}^{(eq)}(y,0)+p_{S}^{(noneq)}(y,t_{2}-t_{1}).
$$
The calculation done in Section \ref{denspict} proves that 
$\mbox{Tr} 
[p_{S}^{(noneq)}(y,t_{2}-t_{1})] = 0$, with the symbol Tr denoting, for simplicity, the integration over $y$ from $0$ to $2$. To 
evaluate the fourth-order correlation function, after applying the 
operator $\xi$ for the second time, we must study the time evolution 
of $p_{S}^{(noneq)}(y,t_{2}-t_{1})$ from $t_{2}$ to $t_{3}$, yielding 
to $p_{S}^{(noneq)}(y,t_{3}, t_{2}, t_{2}-t_{1})$. This is a 
contribution generating some concern, since it activates again the 
back injection process, which we have seen to be intimately related 
to the deviation from equilibrium. However, the compatibility with 
equilibrium condition is ensured by the property $\mbox{Tr}[p_{S}^{(noneq)}(y,t_{3}, t_{2}, t_{2}-t_{1})] = 0$. 

At time $t_{3}$ we have to apply the operator $\xi$ again, and this allows us to make 
an excursion in the anti-symmetric representation, with the time 
evolution given by the operator $\exp(\hat{\Gamma}(t_{4}-t_{3}))$. At time 
$t_{4}$ we apply the operator $\xi$ again, we go back to the 
symmetric representation and we conclude the calculation by means of 
the trace operation. 

In conclusion, the compatibility with equilibrium is guaranteed by the fact that at the intermediate steps of the calculation $\mbox{Tr}[p]= 0$
(the final step, of course, generates  the density-generated correlation function, thereby implying $\mbox{Tr}[p] \neq 0$). 
If the intermediate $p$ is anti-symmetric, this condition is obvious. 
If the intermediate $p$ is symmetric, the vanishing trace condition generates the apparently unphysical property 
that the symmetric contribution gets negative values over some portions of the interval $I$. 
We have to stress, however, that the Liouville-like approach illustrated 
in this paper keeps the distribution density $p(y,t)$ definite 
positive, as it must. The generation of a negative distribution 
density refers to the calculation of the equilibrium correlation 
functions, of any order, and it is a quantum-like property that must 
be adopted to guarantee that the genuine distribution density remains 
in the equilibrium condition.

\section{concluding remarks}\label{concl}
In conclusion, this paper 
shows how to derive the equilibrium correlation function using only 
information afforded  by the Liouville-like approach. The intriguing 
problem to solve was how to use the Liouville equation, without 
conflicting with the equilibrium condition. The solution of this 
intriguing problem is obtained by splitting the Liouville equation 
into two independent components,  the symmetric component 
corresponding to Eq. (\ref{new_evol_symm}) and the anti-symmetric component corresponding to 
Eq. (\ref{new_evol_anti}). This splitting 
allows us to study the regression to equilibrium of the correlation 
function $\Phi_{\xi}(t)$, through Eq. (\ref{new_evol_anti}), without 
ever departing from equilibrium, a physical  condition that is 
controlled by the independent equation of motion for the symmetric 
component, Eq. (\ref{new_evol_symm}). This is very formal, and 
Section \ref{onsregr} makes it possible for us to reconcile it with physical 
intuition. We imagine that the Liouville equation is used to evaluate 
the difference between the population of the left and right state. 
This makes it possible to establish a direct connection between the 
experiment of regression to equilibrium and the formalism of Sections \ref{themodel} and \ref{denspict}. 
We have to create an asymmetrical initial condition, with 
more population on the left than on the right. The time evolution of 
$\Delta P(t)$ depends only on the time evolution of the 
anti-symmetric component of the distribution density, and 
consequently only on the operator $\hat{\Gamma}$. 
This is the reason why 
it is possible in principle to connect regression from an out of 
equilibrium initial condition to the equilibrium correlation 
function, a condition that implies no deviation from equilibrium 
distribution. However, creating in a finite time an out of 
equilibrium condition such that  the left part of the anti-symmetric 
component is identical to the left part of the equilibrium 
distribution, is impossible. This is the reason why we predict the 
breakdown of the Onsager principle in general. 

Section \ref{higherorder} explains why in the literature on dichotomic fluctuations 
the factorization assumption of Eq. (\ref{factorization}) is often 
made regardless of the Poisson or non-Poisson nature of the 
underlying process. See, for instance, the work of Fulinski
\cite{fulinski} as well as Ref. \cite{maurobologna}.  In fact, if 
condition (a) applies, the higher-order correlation functions are 
factorized, thereby making their calculation easy.  However, this 
assumption conflicts with the trajectory arguments of Ref. 
\cite{allegro}, which prove that the factorization condition is 
violated by the non-Poisson condition.  
A rigorous use of the 
Liouville equation shows that condition (a) does not apply, and that 
we have to use condition (b) instead. The calculation of the 
fourth-order correlation function is not straightforward, and this is 
the reason why, to the best of our knowledge, it was never done using 
the Liouville approach. 

This is a fact of some relevance for the 
creation of master equations with memory. There are two major classes 
of generalized master equations. The first class is discussed, for 
instance, in Ref. \cite{gerardo}. The master equations of this class 
are equivalent to the Continuous Time Random Walk (CTRW) of Montroll and Weiss 
\cite{montroll} and are based on the waiting time distribution 
$\psi(t)$. The second class of master equations is based on the 
correlation function $\Phi_{\xi}(t)$. Recent examples of this second 
class can be found in Ref. \cite{maurobologna} and in Ref. 
\cite{pala}. Due to the direct dependence on the correlation function 
$\Phi_{\xi}(t)$, the derivation of the master equations of this 
second class is made easy by the factorization assumption. It must be 
pointed out, on the other hand, that the factorization property, 
which is not legitimate with renewal non-Poisson processes, is a 
correct property if the deviation from the exponential relaxation is 
obtained by time modulation of a Poisson process \cite{pala}. 
Beck 
\cite{beck} is the advocate of the modulation process as generator of 
complexity. Thus, we find that the master equations of the first 
class are generated by the renewal perspective of Montroll and Weiss 
\cite{montroll} and those of the second class are the appropriate 
tool to study complexity along the lines advocated by Beck 
\cite{beck}. We think that the results of the present paper might 
help the investigators in the field of complexity to make the proper 
choice, either modulation or renewal\cite{note}, or a mixture of the two 
conditions.

\emph{Acknowledgements}. PG thankfully aknowledges the Army Research Office for financial support through Grant DAAD19-02-0037


\begin{thebibliography}{99}
\bibitem{note} In the case where the renewal condition applies,  we want to point out that, as shown in \cite{gerardo}, the generalized master equation can be derived from the CTRW \cite{montroll}. It is still unknown how to do that using the Liouville-like approach of Section \ref{themodel}, probably due to the complexity of the problem illustrated in Section \ref{higherorder}. Furthermore, the authors of Ref. \cite{sokolovblumenklafter} made the following impressive discovery. In the non-Poisson case even if a correct generalized master equation is available \cite{note2}, its response to external perturbations departs from the correct prediction that one can obtain by perturbing the  CTRW instead. Thus the trajectory-density conflict \cite{maurobologna} seems to be  the manifestation of  technical problems, emerging from the density perspective, which can be settled using the trajectory (CTRW) perspective.
\bibitem{note2} Within the theoretical framework of this paper, the fractional Fokker-Planck equation used by Sokolov, Blumen and Klafter \cite{sokolovblumenklafter} belong to the class of generalized master equations that can be derived from the CTRW,
\end{thebibliography}
\end{document}